\documentclass[nameyear]{elsart}
\journal{New Astronomy}

\usepackage{epsfig}

\def\astrobj#1{#1}

\newcommand{\msun}{\mbox{$M_\odot$}}
\newcommand{\rsun}{\mbox{$R_\odot$}}
\def\be{\begin{eqnarray}}
\def\ee{\end{eqnarray}}
\def\lsim{\mathrel{\rlap{\lower3pt\hbox{\hskip1pt$\sim$}}
     \raise1pt\hbox{$<$}}} %less than or approx. symbol
\def\gsim{\mathrel{\rlap{\lower3pt\hbox{\hskip1pt$\sim$}}
     \raise1pt\hbox{$>$}}} %greater than or approx. symbol

\begin{document}

\runauthor{Brown \& Lee}

\begin{frontmatter}
\title{The Case for Case C Mass Transfer in the Galactic
       Evolution of Black Hole Binaries}

\author[suny]{Gerald E. Brown\thanksref{geb}}
\author[pnu]{Chang-Hwan Lee\thanksref{chlee}}

\address[suny]{Department of Physics and Astronomy,\\ 
State University of New York, Stony Brook, NY 11794-3800}

\address[pnu]{
Department of Physics and\\
Nuclear Physics \& Radiation Technology Institute (NuRI), \\
Pusan National University, Busan 609-735, Korea\\
  }
\thanks[geb]{Ellen.Popenoe@sunysb.edu}
\thanks[chlee]{clee@pusan.ac.kr}

\begin{abstract}
Earlier works, which we review, have shown that if the Fe core in a
presupernova star is to be sufficiently massive to collapse into a 
black hole, earlier in the evolution of the star the He core must be
covered (clothed) by a hydrogen envelope during He core burning and
removed only following this, in, e.g. common envelope evolution.
This is classified as Case C mass transfer. These previous arguments
were based chiefly on stellar evolution, especially depending on the
way in which $^{12}$C burned.

In this work we argue for Case C mass transfer on the basis of binary
evolution. The giant progenitor of the black hole will have a large
radius $\sim 1000\rsun$ at the end of its supergiant stage. Its 
lifetime at that point will be short, $\sim 1000$ yrs, so it will
not expand much further. Thus, the initial giant radius for Case C
mass transfer will be constrained to a narrow band about $\sim 1000\rsun$.
This has the consequence that the final separation $a_f$ following
common envelope evolution will depend nearly linearly on the mass
of the companion $m_d$ which becomes the donor after the He core
of the giant has collapsed into the black hole. The separation
at which this collapse takes place is essentially $a_f$, because of
the rapid evolution of the giant. (In at least two binaries the black
hole donor separation has been substantially increased because of mass loss
in the black hole formation. These can be reconstructed from the 
amount of mass deposited on the donor in this mass loss.)

We show that the reconstructed preexplosion separations of the black hole
binaries fit well the linear relationship.
\end{abstract}


\end{frontmatter}

\def\COrate{$^{12}$C($\alpha,\gamma$)$^{16}$O } 

%%%%%%%%%%%%%%%%%%%%%%%%%%%%%%%%%%%%%%%%%%%%%%%%%%%%%%%%%%%%%%%%%%%%%%%%%%%%%%%%
\section{Introduction\label{intro}}

The black hole Soft X-ray Transient (SXT) \astrobj{A0620$-$00} consisting
of an $\sim 11\msun$ black hole and $\sim 0.7\msun$ K-star companion
was evolved theoretically by De Kool et al. (1987) with Case C mass
transfer; i.e., the common envelope evolution in which the companion
removed the hydrogen envelope of the giant black hole progenitor
through spiral-in took place after He
core burning was completed in the giant. 
In Table~1 of Lee, Brown, \& Wijers (2002) (denoted as LBW), 
the  seven SXTs with shortest periods had K- or M-star
companions and the unclassified companion in \astrobj{XTE 1859$+$226} may also
well be K or M because of its short period $P=0.380$ days. 
The progenitor binaries of these would have involved
$\sim 25\msun$ giants and $\sim 1-2\msun$ companions, the latter having
had some mass stripped off by the black hole.
In other words, all of the shortest period SXTs are successfully evolved
with the {\it same} Case C mass transfer. 

We emphasized in Lee \& Brown (2002)
that in Case C mass transfer the orbital separations for the 
above progenitor binaries in Roche Lobe contact are
at $\sim 1700\rsun (\pm \sim 10\%)$, $\sim 8$ AU, for ZAMS $20\msun$
black hole progenitor and $1\msun$ companion star.
Progenitors with more massive
companions and the larger initial separation 
necessary for Case C mass transfer could have removed the H-envelope
of the giant with spiral-in to larger final separations $a_f$, since
their drop in gravitational energy of the more massive companion is then
sufficient to remove the envelope, and we shall see that this is indeed
what happens.

In LBW we listed \astrobj{Nova Scorpii}  and \astrobj{IL Lupi} 
as undergoing mass transfer
while in main sequence. Beer \& Podsiadlowski (2002) have carried out
a detailed, convincing numerical evolution of \astrobj{Nova Scorpii}, showing
that the orbit has widened substantially under nearly conservative
mass transfer. Podsiadlowski et al. (2002) (denoted as PRH) have recently
extended such calculations to the other binaries with evolved
companions, showing that they all began mass transfer in main
sequence, although \astrobj{V404 Cyg}, \astrobj{J1550$-$564} 
and probably \astrobj{GRS 1915$+$105} will have
progressed beyond main sequence. (As noted later, we differ
with PRH in our suggested evolution of
\astrobj{Cyg X-1}.) The PRH calculations
generally support the schematic LBW calculations of mass transfer,
but have the added advantage that by beginning the transfer in main
sequence, sufficient mass can be transferred in the traditional
sub Eddington limit. Whereas we do not believe this to be
necessary in the case of black holes, seeing no reason why
the accretion across the event horizon could not be substantially
hyper Eddington (and PRH also covers
this case, as a possibility) the standard PRH
scenario allays the fear of the greatly hyper Eddington scenario
which may go against "accepted wisdom".

In Case C mass transfer there is a great regularity expressed in the
roughly linear
dependence of companion mass on orbital separation of the giant
black hole progenitor and companion on the companion mass
\be
a_f \propto \frac{M_d}{\msun} \left(\frac{M_{\rm giant}}{\msun}
\right)^{-0.55} R
\label{eq1}
\ee
following the spiral-in stage which removes the envelope of the
giant (Lee, Brown \& Wijers 2002). 
We derive Eq.~(\ref{eq1}) in the Appendix.
Except for the roughly square root dependence on giant mass,
this relation is linear.
Here the companion (donor)
mass is labelled $M_d$, $a_f$ is the separation of the He-star,
companion binary following spiral-in in common envelope evolution,
and $R$ is the initial radius of the giant at the start of common envelope
evolution. The dependence on $M_{\rm giant}$
is weak, the interval
\be
20\msun < M_{\rm giant} < 30\msun
\ee
being used by Lee, Brown \& Wijers (2002). The term depending
on giant mass originates from the term $M_{\rm He}/M_{\rm giant}^2$
in common envelope evolution. 
The relation Eq.~(\ref{eq1}) is particularly useful because, as we
shall argue, $R$ is nearly constant, $\sim 1000\rsun$, to within
$\sim 10\%$. Because the giant evolutionary time is so short, $a_f$
is essentially the preexplosion separation of black hole and donor.

We thus have three classes:

\begin{enumerate}

\item[(i)] The 8 AML (angular momentum loss) SXTs with K or M-star 
main sequence companions come from binaries
which overfill their Roche Lobe during spiral in, as discussed in LBW.
Their periods are decreased as they transfer mass to the companion
black hole, as they lose angular momentum by magnetic braking and
gravitational waves.

\item[(ii)] The next six SXTs which established Roche contact while
in main sequence, some of them having evolved beyond.

\item[(iii)] The special case of the continuously shining
\astrobj{Cyg X-1} which we place just before its Roche Lobe, the
companion now undergoing unstable mass transfer to its lower mass
companion black hole.

\end{enumerate}

Interestingly, we find that the division between the unevolved main sequence
class (i) and evolved companion (ii) is given accurately by Fig.~2
of de Kool et al., who plot the mass of the companion which undergoes
angular momentum loss by gravitational waves and magnetic braking,
both as functions of time. They obtain the companion mass of $2\msun$
as giving the division. We find this to be true for $M_{\rm giant}
=20\msun$ in Eq.~(\ref{eq1}). 

Note that the binaries with late main sequence companions \astrobj{Nova Scorpii}
and \astrobj{IL Lupi} are special in that these binaries
have experienced large mass loss, which can be explained as in LBW
by magnetohydrodynamic effects, not included in the evolution discussed here.
One may wonder why just there two binaries, with ZAMS companion masses,
have lost a sizable fraction of their progenitor He star masses,
whereas there is no sign of a kick outwards in separation from copious
mass loss in the Class (i) SXTs.

The above regularity shows immediately that at least the first class
with K and M companions must have a very large $a_i$ (LBW find
$a_i\sim 1700\rsun$ for a $20\msun$ black hole progenitor with $1\msun$
companion) corresponding to a giant radius of $\sim 1000\rsun$, 
so that the binding energy of the giant envelope, which
decreases inversely with its radius, is small enough to be furnished
by the drop in gravitational binding energy of the low-mass companion
as it spirals in to its Roche Lobe.

As developed in many papers by Brown and collaborators, most lately
in Lee \& Brown (2002) and LBW, and backed by evolutionary
calculations by Brown et al. (2001) and PRH, the above delineation into three
classes, depending upon companion mass, can be understood if the giant
is required to finish (or nearly finish) He core burning before common
envelope evolution takes place; i.e., if the mass transfer is
essentially Case C.

In Sec.~\ref{caseC}, 
we discuss the stellar evolution necessary to produce
models which allow Case C mass transfer for ZAMS $20-30\msun$ stars.
In Sec.~\ref{sec_dev}, we review the role of carbon burning.
In Sec.~\ref{popul}, we discuss that Case B mass transfer
would not only allow too much of the He envelope to blow away
and leave too much $^{12}$C after He core burns, but also 
is disfavored by the population of SXTs.
In Sec.~\ref{sec_pop},
we compare our approach with other works, especially those by PRH,
and discuss population synthesis of Case B mass transfer.
We summarize our conclusion in Sec.~\ref{disc}. 

%-----------------------------------------------------------------------

\section{The Case for Case C Mass Transfer}
\label{caseC}

In LBW we found that the Schaller et al. $20\msun$ star
had the characteristics we desire for Case C mass transfer, but that the
latter was not possible for their $25\msun$ star. We therefore constructed
``by hand" models in which the stellar radius as function of burning stage
had a similar shape to the $20\msun$ star, all the way up to $30\msun$.

\begin{figure}
\centerline{\epsfig{file=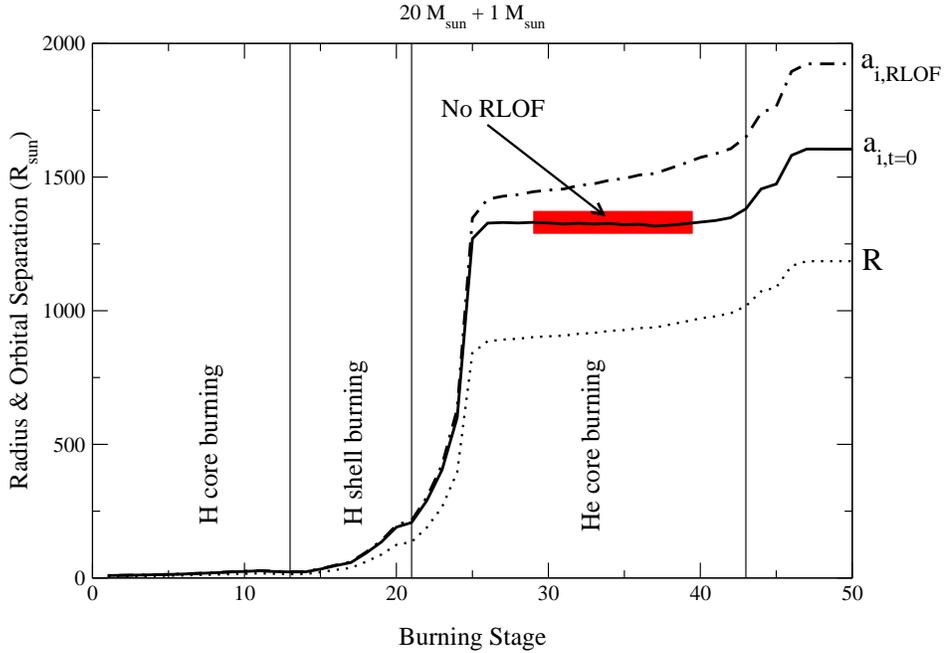,height=4.0in}}
\caption{
Radius of black hole progenitors ($R$) and the
initial orbital separations ($a_i$) of the progenitors of X-ray transient
binaries with a $1\msun$ companion. The burning stage in the x-axis corresponds
to that of Schaller et al. (1992).
{\bf A)} The lower dotted curves ($R$) corresponds to the radius of
the black hole progenitors taken from Schaller et al. (1992).
That for the
$25\msun$ star is similar but for the $30\msun$ the radius does not
increase following the end of He core burning.
{\bf B)}
From the mass of the primary at the tabulated point
one can calculate the semimajor axis
of a binary with a $1\msun$
secondary in which the primary fills its Roche Lobe, and this
semimajor axis  is shown in the upper dot-dashed curve ($a_{i,{\rm RLOF}}$).
{\bf C)} The solid curves ($a_{i,t=0}$) correspond to the required initial
separations after corrections of the orbit widening due to the
wind mass loss, $a_{i,t=0}= a_{i,{\rm RLOF}}\times (M_p+M_d)/(M_{p,0}+M_d)$
where $M_p$ is the mass of the black hole progenitor at a given stage
and $M_{p,0}=20\msun$ is the ZAMS mass of the black hole progenitor.
Primaries at the evolutionary stages marked by the shaded area
cannot fill their Roche Lobe for the first time at that stage,
but have reached their Roche Lobe at an earlier point in their
evolution.
}

\label{bltf1}
\end{figure}

We show in Fig.~\ref{bltf1} the results of the Schaller
et al. (1992) stellar evolution for a ZAMS $20\msun$ star.
It is seen that the main increase in radius comes after the start
of He core burning (which begins while H shell burning is still going on).
With further He core burning there is a flattening off of the radius
versus burning stage and then a further increase in radius towards
the end of and following He core burning. Our model requires that
mass transfer take place during this last period of increase in radius,
so that the orbital separation
($\sim 3/2$ of the giant radius) is well
localized $a_{i,{\rm RLOF}} \sim 1700\rsun$ at the time of Roche Lobe
contact, or $a_{i,t=0}\sim 1500\rsun$ initially, the difference
due to mass loss by wind, with accompanying widening of the orbit.

It is made clear in LBW and Lee \& Brown (2002) that for
Case C (or very late Case B) the radii of the relevant stars must have
the following behavior, as shown in Fig.~\ref{bltf1}.

\begin{enumerate}

\item[(i)] They must increase rapidly in radius
with hydrogen shell burning and with
the early He core burning, which begins while the hydrogen shell
burning is still going on.

\item[(ii)] The radii must flatten off, or actually decrease with further
He core burning. This is so that if the companion reaches the Roche
Lobe it will reach it before or early in He core burning. Then 
according to Brown et al. (2001a) the He core made naked by common
envelope will mostly blow away by the strong Wolf-Rayet type winds,
and the final Fe core will be too low in mass to collapse into
a high mass black hole.

\item[(iii)] The third (obvious) characteristic is that the stellar radius
must grow following He core burning, because the massive star must
be able to reach its Roche Lobe during this time. The massive star
has only $\lsim 10^4$ years of its life left, so wind losses no longer
can carry much of it away.

\end{enumerate}

\begin{figure}
\begin{center}
\epsfig{file=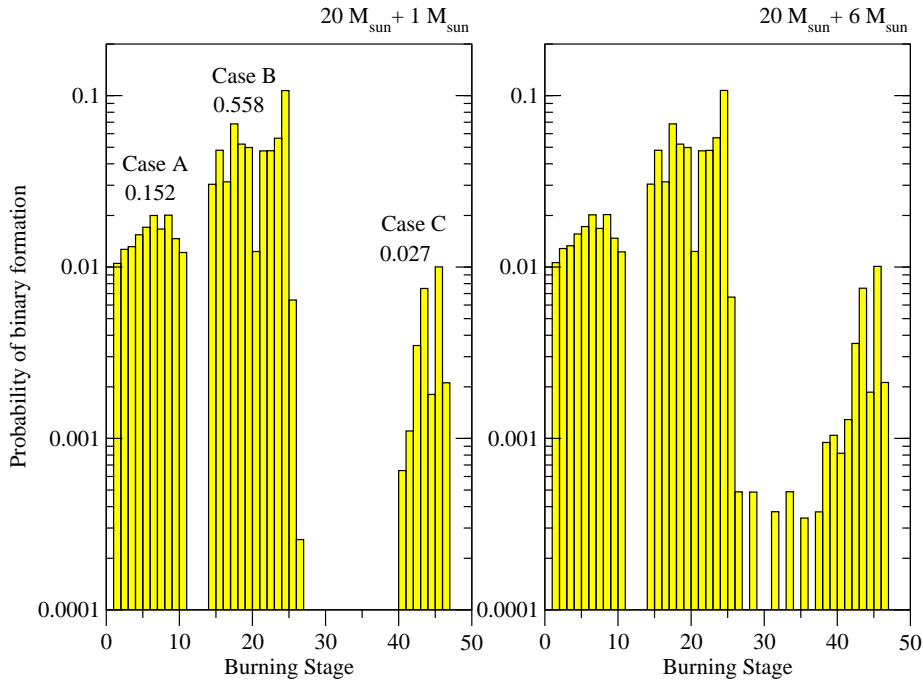,width=1.0\textwidth}
\end{center}
\caption{
Probability of initial binary formation, in which the Roche Lobe overflow
starts between the two adjacent burning stages of the $20\msun$ ZAMS star.
The burning stages are the same as in Fig.~\ref{bltf1}.
The probability (logarithmic distribution of initial binary separation)
is given by $P=\log (a_{n+1}/a_n) /7$ 
where the initial binary separation $a_{i,t=0}$ is between $a_n$ and $a_{n+1}$,
and the logarithmic distribution is normalized by the total 
logarithmic interval ``7" of Bethe \& Brown (1998).
Three different cases of mass transfer are marked by Case A, B, and C.
The numbers for each case in the left panel
are the total of the probabilities in each case. 
In the left hand panel,
the radii for the Case B mass transfer between stage 14 and stage 27 are
$R=22 - 892 \rsun$ with the corresponding initial binary separation 
$a_{i,t=0}=33 - 1330\rsun$.
For Case C mass transfer between stages 40-47, 
$R=971-1185\rsun$ and $a_{i,t=0}=
1331-1605\rsun$.
With a $6\msun$ companion,
the intermediate Case B mass transfer is possible as in the right panel.
However, the total probability for the intermediate Case B mass transfer
is $\sim 10\%$ of that for the late Case B and Case C mass transfer.
 \label{fig2}}
\end{figure}

Portegies Zwart et al. (1997) have pointed out that wind
loss from the giant preceding common envelope evolution is
important and we follow their development in identifying
the ``No RLOF" part of the curve in Fig.~\ref{bltf1}.
Because of the wind loss the binary widens. The Roche
Lobe overflow will take place during the very rapid increase in
radius of the giant in the beginning of He core burning, or in very
late Case B or in Case C mass transfer. The binary has widened
too much by the time the giant has reached the flat part of the
$R$ vs stage curve. In fact, it cannot transfer mass during this
stage, because it will have already come to Roche contact during
early Case B. This is made clear by the shaded area in the
solid line in Fig.~\ref{bltf1}. Brown et al. (2001c) have
shown that the SXTs with main sequence companions can be
evolved with a $1-1.25\msun$ companion mass, so the above
results may be directly applicable.
For higher mass companions, this shaded area becomes smaller
because the effect of the winds is smaller. This
makes the intermediate Case B mass transfer possible.
However, in this case, high-mass black holes may not form
because the Fe core is not massive enough to form high-mass compact
objects as we discussed above (Brown et al. 2001a).
Furthermore, as in Fig.~\ref{fig2}, the probability of the intermediate
Case B mass transfer is small compared to that of Case C mass transfer.

Now, in fact, the curve of radius vs burning stage for the next
massive star, of ZAMS $25\msun$, by Schaller et al. (1992)
does not permit Roche Lobe contact during Case C at all, the winds
having widened the binary too much by the time the giant radius begins
its last increase in late He core burning.
In the ZAMS $30\msun$ star of Schaller et al. (2002), there is no increase
in $R$ at this stage, so Case C mass transfer is not possible.

The lack of increase in $R$ for the more massive stars is due to the
cooling effect by strong wind losses. As shown by Lee et al. (2002)
giant progenitors as massive as $30\msun$ are necessary as progenitors
of some of the black holes in the SXTs, especially for the binaries
with evolved companions, in order to furnish the high mass black hole
masses. These authors reduce wind losses by hand, forcing the resulting
curve of $R$ vs burning stage to look like that for a ZAMS $20\msun$ shown
in Fig.~\ref{bltf1} during the He core burning where the effect
of wind loss is important. In other words, in order to get the observed
regularities in the evolution of SXTs, especially Eq.~(\ref{eq1}) which
gives the linear dependence on $M_d$ of the preexplosion separation
of the binary, we must manufacture $R$ vs burning stage curves for which mass
transfer can be possible both early in Case B and
in Case C. With early
Case B mass transfer, or intermediate Case B mass transfer if it occurs,
the winds during He core burning are so strong
that not enough of an Fe core is left to result in a high-mass black hole,
rather, a low-mass compact object results (Brown et al. 2001a).

This story is somewhat complicated, but there have been many years of
failures in trying to evolve black holes in binaries without taking
into account the effects of binarity (mass transfer in our model)
on the evolution. On the other hand, de Kool et al. (1987)
had no difficulty in evolving \astrobj{A0620$-$00} in Case C mass transfer. The
necessity in a similar evolution for the other black hole
binaries was, however, not realized at that time.

%%%%%%%%%%%%%%%%%%%%%%%%%%%%%%%%%%%%%%%%%%%%%%%%%%%%%%%%%%%%%%%%%%%%%%%%%%
\section{Dependence on the {\boldmath \COrate} Rate}
\label{sec_dev}

Brown et al. (2001a) showed that the mass at which single stars
went into high-mass black holes was determined by the 
$^{12}$C($\alpha,\gamma$)$^{16}$O rate. For the Woosley rate of 170 keV barns,
stars from $8-18\msun$ would go into neutron stars, the narrow range from
$18-20\msun$ into low-mass black holes (We believe 1987A to be an example.)
and the stars from $20\msun$ on up to a maximum mass determined by wind losses,
possibly $\sim 30\msun$ into high-mass black holes.

The main conclusion of Brown et al. (2001a) was that the massive star
must be clothed by its H envelope during most, if not all, of its He
core burning, if the core is to be massive enough so as to collapse
into a high-mass black hole.

Schaller et al. used $\sim 100$ keV barns for the \COrate rate, and
we can check that their central $^{12}$C abundance following He
core burning goes down to $\sim 15\%$ for their $25\msun$ star.
In fact their $25\msun$ star does expand quite rapidly just at their
stage 43, the end of He core burn. However, large wind losses cause
the binary to widen too much for Case C mass transfer, and these
must be cut down somewhat as done by LBW if Case C
is to be made possible.

One consequence of the skipping of convective carbon burning is
that the remaining lifetime of the core should be substantially
foreshortened. Whereas convective carbon burning takes hundreds
of years, neon and oxygen burning take only $\sim$ one year.
The interpolation from $^{12}$C to $^{16}$O burning via
radiative and shell $^{12}$C burning and neon burning, which remains
even when the central $^{12}$C is less than $15\%$, 
will smooth out any abrupt change, but the foreshortening
should none the less be appreciable.
It lessens the time available for tidal interactions in the
He-star, donor binary lifetime.

Our considerations apply to Galactic metallicity. With low metallicity,
the opacity is less and winds would not be expected to blow off
naked He envelopes. Thus, Case A, AB or B mass transfer might not
be expected to lead to only low-mass compact objects. 
The LMC with metallicity about 1/4 Galactic, has two continuously
shinning X-ray binaries, LMC X-1 and LMC X-3, even though the total
LMC mass is only $\sim 1/20$ of Galactic.

There is an important caveat to the large expected effect from
lower metallicity and stronger winds. As discussed in Brown et al. (2001a)
(see their Table~2) the mass loss rate has to be lowered by a factor
of 3 from the preferred rate
(which fits the fractional period
change $\dot P/P$ in \astrobj{V444 Cyg}) before the convective $^{12}$C
burning is skipped (with a central 12\% $^{12}$C abundance).
In fact, even then (Fryer et al. 2002)
the compact core is only 1.497, only large enough to collapse into
a low-mass compact object. But, in the Fryer et al. (2002) calculations,
the compact core is brought back up to $10.7\msun$ by fallback,
sufficient for collapse into a high-mass black hole (as in the
$5.2\msun$ remnant obtained when the mass loss rate is cut down
by a factor of 2, rather than 3). 
However, (Brown et al. 2000; LBW) in a binary magnetohydrodynamic
effects should help expel the outer matter in the explosion,
cutting down the fallback.

From the above one can see that even cutting winds down will not
necessarily make Case B mass transfer possible. Perhaps more
important is the lack of $^4$He needed to burn the last
$^{12}$C left. As the triple alpha reaction depends on the
third power of the helium mass fraction it loses against the
\COrate reaction toward the end of central helium burning; i.e.,
carbon is mostly burned rather than produced toward the end
of central helium burning. That switch typically appears at a
central helium mass fraction of $\sim 10-20\%$. Most importantly,
as can be seen from the central carbon abundances at the end
of He burning, which decreases from 35\% to 22\% with the
lowering of wind losses by a factor of 6 (Fryer et al. 2002)
the He fraction is too low to burn the final carbon.
Only with the 6-fold reduction in wind from the Woosley, Langer, \&
Weaver (1995) rate (which is 3-fold from our preferred value)\footnote{ 
    Half of the Woosley, Langer \& Weaver (1995) rate.  }
is convective
carbon burning skipped. (With a 4-fold lowering from WLW, the convective
carbon burning goes on for 500 years.)

In the clothed stars, on the other hands, the growth of the
He core and accompanied injection of helium after this
time leads to a further decrease of carbon as compared
to the bare helium cores that do not have this additional
supply of helium. We believe the above may be the most
important difference between naked and clothed He cores.

%%%%%%%%%%%%%%%%%%%%%%%%%%%%%%%%%%%%%%%%%%%%%%%%%%%%%%%%%%%%%%%%%%%%%%%%%%

%------------------------------------------------------------------
\section{Evolutionary Consequences If Case B Were Possible}
\label{popul}

\begin{figure}
\centerline{\epsfig{file=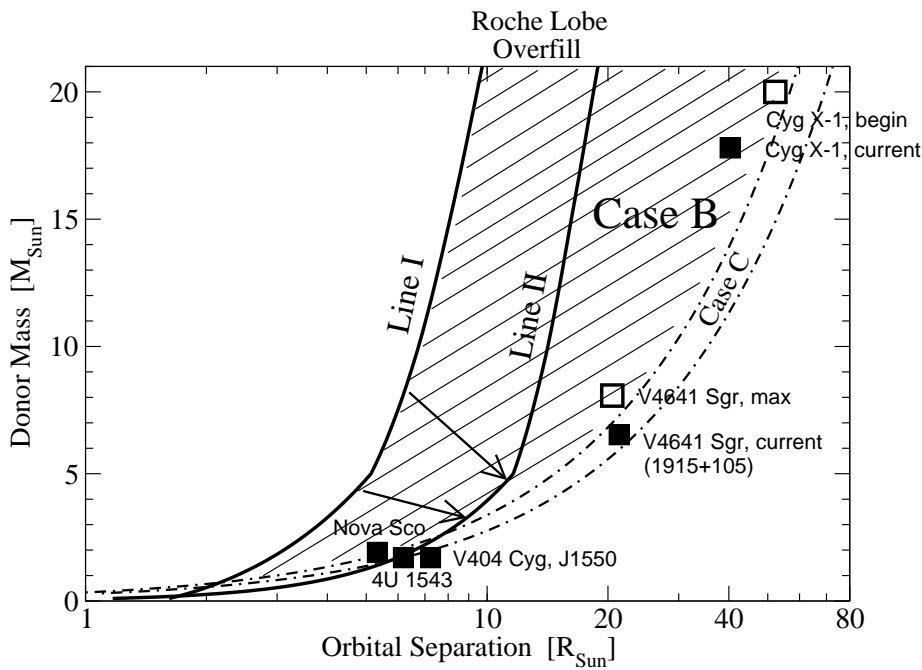,height=4in}}
\caption{Orbital separations after common envelope evolution
for Case B and Case C mass transfer. 
Dot-dashed lines are the limits for the Case C mass transfer with
$M_p=30\msun$, $M_{\rm He}=11\msun$, and $\lambda\alpha_{ce}=0.2$
(see LBW). 
All the area to the left of the left dot-dashed line is Case B
mass transfer (shaded area).
Line I is the sum of the radius of the companion and
that of the He core
which is assumed to be $1.5\rsun$. Line II is the orbital
separation corresponding to the Roche Lobe overfill right after
the spiral-in during the common envelope evolution. The companion
stars in binaries between Line I and II will be inside their Roche Lobe
(Roche Lobe overfill) when they finish  common envelope evolution,
and they will be pushed out with mass transfer as indicated by arrows
or they will lose in the common envelope evolution (see Fig.~\ref{fig4}).
Those binaries ($M_{\rm donor}>2.5\msun$)
between Line II and the left boundary of Case C
will be outside of the common envelope even with Case B mass transfer.
Reconstructed preexplosion orbital separation and black holes masses
of SXTs with evolved companions are marked by black squares (refer to
Fig.~11 of LBW.) If the high mass black hole
formation in Case B mass transfer were possible, the probability
of observing them in Case B is $\sim 7$ times larger than in Case C.
However, for the donor masses $\gsim 2\msun$, we see no SXTs in Case B,
while we have two observations, \astrobj{V4641 Sgr} and \astrobj{GRS 1915$+$105}
 in
which the reconstructed data is consistent with Case C.
We have put in both the reconstructed data with maximum initial black
hole mass (open square), and the present position of \astrobj{V4641 Sgr}
(filled square)
in order to show
the uncertainty in reconstruction. The small change in orbital
separation shows this binary to give an excellent fiducial 
preexplosion separation. Because of the long period, mass loss
in the explosion will be low (LBW).
\astrobj{Cyg X-1} 
may have had the preexplosion separation shown by the open box;
its current separation is shown by the filled box.
Although the $a_f$ is linear 
with companion mass $M_d$,
the curves delineating Case C mass transfer curve up when the
orbital separation is plotted logarithmically.
}
\label{fig3}
\end{figure}

We see from Eq.~(\ref{eq1}), or equivalently, Eq.~(\ref{eqA3})
that the preexplosion separation $a_f$ scales linearly with $M_D$,
a relation that was used in LBW to evolve all binaries with evolved
companions. Note that the range of $a_i$ also depend on the donor masses 
through the changes in Roche Lobe radii as 
in Fig.~\ref{fig3}.
In LBW we used this scaling, which also followed from
the Webbink common envelope evolution, and showed that the evolution
of all of the SXTs could be understood in terms of it.

During H shell burning and He core burning the radius $R$ of the
giant increases rapidly up to $\sim 892\rsun$. (During the increase
from $\sim 892\rsun$ to $971\rsun$ the wind losses widen the orbit
at such a rate that there is no RLOF as shown in Fig.~\ref{bltf1}.
The decrease in the $a_{i,t=0}$ curve is even more pronounced
in the curve of Heger (private communication, 2000). This may
change with decreased wind losses, but we expect any increase
in the $a_{i,t=0}$ to be small, and neglect it here. Consequently,
Case B mass transfer could be early, taking place with H shell
burning or early He core burning.

Suppose Case B mass transfer takes place during the stages between
15 and 27 as in Fig.~\ref{fig2}. Certainly it can, although
we say the results will be a binary with a low-mass compact object. 
It would most likely do so for radii from $\sim 22$ to $892\rsun$,
and the corresponding initial binary separation from $\sim 33\rsun$
to $1330\rsun$.
We set the lower limit to be the radii at the stage 15 following
the gap between Case A and Case B mass transfer in Fig.~\ref{fig2}. 
For the
total binary logarithmic interval we take the 7 of Bethe \& Brown (1998).
With the above $33-1330\rsun$ the fractional logarithmic interval is
$\ln(1330/33)/7\sim 0.53$ whereas for Case C mass transfer it is
$\ln (1604/1331)/7=0.026$. Thus, for a logarithmic distribution of binaries,
Case B mass transfer is favored by a factor $\sim 20$. What would
the consequences of this be, assuming it to be possible ?

First of all let us consider SXTs like \astrobj{V4641 Sgr} which is just 
beginning to cross
the Herzsprung gap; this consideration also includes \astrobj{GRS 1915$+$105}
which was shown by LBW to be a late \astrobj{V4641 Sgr}
on the other side of the
Herzsprung gap. Both had the large,
ZAMS $\sim 6.5-8\msun$ companions. 
\astrobj{V4641 Sgr} has at present radius $R=21.3\rsun$.
and from
the closeness of black hole and companion masses, could not have been
narrower than $\sim 20.5\rsun$ ($=21.3\times 
(\frac{9.61\times 6.53}{8.07^2})^2\rsun$)
at which separation the companion and
black hole mass would have been of nearly equal mass,
which could have been as massive as $8\msun$ originally.

By way of example of how Case B mass transfer might function,
we consider binaries with $M_{\rm He}=11\msun$ and
$M_{\rm D}=8\msun$; i.e., binaries similar to our reconstructed 
\astrobj{V4641 Sgr}
at the time of black hole formation, as an example. 
The orbital separation for Roche Lobe overflow 
is $\sim 13.2\rsun$, taking the donor
radius to be $4.5\rsun$. 
The radius of an $11\msun$ He star is
$1.5\rsun$, so the sum of donor radius plus He-star radius is
$6\rsun$. If the binary separation is smaller than this, it will
merge during the evolution. So the range of orbital separations for Roche
Lobe overfill after common envelope evolution is $\sim 6-13\rsun$.
This means that if the companion star spirals in
from anywhere between an initial $823$ and $1770\rsun$ it
will overfill its Roche Lobe.\footnote{
   The initial orbital separation for Case C mass transfer
   is larger than those in Fig.~\ref{fig2} due to the larger
   radius of the companion star.}
It will then transfer mass to the
He star until it fits into its new Roche Lobe with reduced mass.
\footnote{The He star may accept some of the mass or it may be lost
in common envelope evolution.  }
Because of the substantial logarithmic interval $\ln (1770/823)/7=0.11$,
nearly 1/5 of the entire Case B logarithmic
interval, or nearly 4 times the entire Case C logarithmic interval.
Conditions of Roche Lobe overfill for various donor masses are
summarized in Fig.~\ref{fig3}.

\begin{figure}
\centerline{\epsfig{file=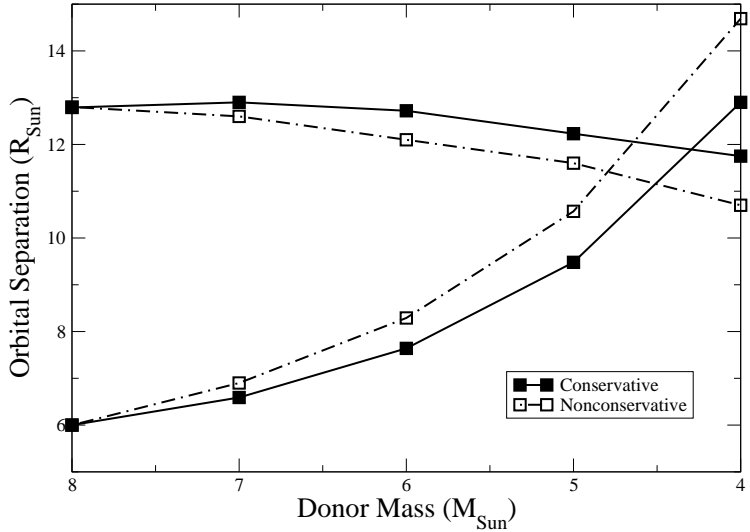,height=3.5in}}
\caption{
Evolution of the orbital separations of the binary which overfill
the Roche Lobe after common envelope evolution.
We assumed the binary separation after common envelope evolution
$a_f=6\rsun$ (left end point of lower curve), 
the sum of the companion and the He core radii
(Line I in Fig.~\ref{fig3}),
with $M_{\rm He}=11\msun$ and $M_{\rm D}=8\msun$.
Since the donor is inside its Roche Lobe, the outer envelope
will be transferred to the He core until the donor fits its
Roche Lobe.
The lower curves correspond to the orbital separations
during the mass transfer. 
The upper curves correspond to the outer radius for the Roche Lobe
overfill (Line II in Fig.~\ref{fig3}) 
with corresponding donor and He core masses.
Roche Lobe overfill will occur for any $a_f$ between $6\rsun$ and the
upper curve, and there will be additional lower curves corresponding
to these different $a_f$.
The mass transfer will continue until the lower curve reaches
the upper one. Once the lower curve reaches the upper one, there
is no further mass transfer and the orbital evolution will stop.
In the conservative mass transfer, we assume
that 100\% of the mass lost from the donor is accreted onto the He star,
and in the nonconservative mass transfer case, we assumed that
all the transferred mass from the donor is lost.
(It may be expelled in the common envelope evolution.)
In both cases, the mass transfer will stop before the donor reaches 
$\sim 4\msun$
where the Roche Lobe is larger than the donor radius.
In the same way, we will arrive $M_d \sim 3\msun$ ($1.5\msun$) if we start
with $M_d=4\msun$ ($2\msun$) as indicated in Fig.~\ref{fig3}.
}
\label{fig4}
\end{figure}

Now consider the case that the binary ended up with the orbital
separation of $6\rsun$ after common envelope evolution.
Since the separation for the Roche Lobe filling is
$4.5\rsun/0.35\sim 12.9\rsun$, the donor
will lose mass until the radius of the donor fits
its Roche Lobe. If we assume  conservative mass transfer, the
mass transfer will stop before the donor reaches $\sim 4\msun$
as in Fig.~\ref{fig4}.
Furthermore, if the explosion occurs before the donor radius
fits its Roche Lobe, the orbit will be widened during the explosion. 
This will reduce the mass loss from the donor star,
so the final reduced main sequence mass will be larger than $4\msun$
as indicated by arrow in Fig.~\ref{fig3}.
%-------------------------------------------------

The chance of seeing the SXTs with massive companions before they
evolve
is small, even if the high mass black hole
formation is possible in Case B mass transfer, either because the
binary will end up beyond its Roche Lobe on Line II or because
the mass transfer  during the early main sequence stage will be  small
if it ended up along the Line II and will increase the binary
radius beyond the Roche Lobe.
Note that the life time of main
sequence with ZAMS mass $\gsim 2\msun$ is shorter than the time
scale of the orbital separation (e.g., Fig.~2 of De Kool et al. 1987, without
magnetic braking).
Instead, they will become SXTs when they evolve.
In the case of \astrobj{V4641 Sgr}
with initial companion mass $M_d=8\msun$, 
$M_{\rm BH}=8\msun$ we estimate that after common envelope evolution
$a_f\sim 1.5 R_L$. Since the radius more than doubles in late main sequence 
evolution (Schaller et al. 1992)
it will reach its Roche Lobe before then. We thus find that
all companions with masses $>2\msun$, aside from that in \astrobj{Cyg X-1},
establish Roche contact in main sequence.
In Fig.~\ref{fig3}, therefore, all binaries between Line I and the boundary
of Case B and Case C will become SXTs with evolved companions,
if high mass black hole formation
in Case B mass transfer were possible. 
In that case, from Fig.~\ref{fig3}, one can see that there should 
be $\sim 8$ times
more SXTs with evolved companions (with initial donor mass $>2.5\msun$).
On the other hand, from Fig.~\ref{fig3},
we expect $\sim 4$ times more SXTs with companions in main sequence
if Case B mass transfer were possible.

Chiefly we see from our discussion of possible Case B mass transfer
in the SXT evolution that there would be no correlation between
companion mass and preexplosion separation, since the possible
initial separations $a_i$ would be very widely spread.
In many cases, the orbit following spiral-in would overfill its Roche
Lobe and mass exchange or loss would spread out the companion masses,
each binary filling its Roche Lobe.
The validity of Eq.~(\ref{eq1}) depends on the possible
post supergiant radii $R$ being within a narrow range,
consequently a narrow range in the preexplosion orbital separation
$a_i$.
We show in Fig.~\ref{fig3} that empirically the relation Eq.~(\ref{eq1})
is satisfied with our preferred common envelope efficiency
$\lambda\alpha_{ce}=0.2$ of LBW.
Of course this depends on the reconstruction of
preexplosion orbits by LBW, which generally is supported by PRH
although they give a wide range of possibilities.

LBW noted that the evolution of \astrobj{Cyg X-1} also fits into our Case C mass
transfer scenario as in Fig.~\ref{fig3}. 
Assume the progenitor of the black hole to be
a ZAMS $25\msun$ giant (with $8.5\msun$ He star which we assume to go
into a black hole of the same mass because very little mass is lost
in the case of such a long period\footnote{
A correction may have to be put in for mass loss in the explosion
forming the black hole (Kaper et al. 1999) although the system
velocity may be only $\sim 1/3$ the 50 km s$^{-1}$ found there,
depending on the O-star association (L. Kaper 2001, private communication).}
).
Following the supergiant stage of the massive giant a ZAMS $20\msun$
companion removes the envelope, coming to an $a_f$ of $\sim 50\rsun$
($=(20\msun /8\msun )\times 20\rsun$),
where $8\msun$ and $20\rsun$ are the reconstructed black hole mass and
$a_f$ at the time of explosion in \astrobj{V4641 Sgr}. We followed the linear
scaling of $a_f$ with $M_d$ here. The companion now transfers
$2.2\msun$ to the black hole in unstable, but conservative mass
transfer. This brings the separation $a$ down to the present $40\rsun$.

%------------------------------------------------------------------
\section{Comparison with Other Works and Population Synthesis}
\label{sec_pop}

As noted earlier, a comprehensive numerical evolutionary calculation
has been carried out by Podsiadlowski et al. (2002; PRH), 
who also adopt
case C mass transfer. Out approach as that in LBW is schematic, but
there is substantial agreement between PRH 
and LBW on most aspects of mass transfer and the effects that follow
from it. This is not surprising since LBW built their work on
the earlier evolutionary calculation for \astrobj{Nova Scorpii} of
Beer \& Podsiadlowski (2002).

PRH agree with LBW that present evolutionary calculations for giant
ZAMS masses $>20\msun$ do not allow Case C mass transfer, and that
these must be changed.

The LBW approch was to cut down on wind loss so as to make
giants from ZAMS $20-30\msun$ behave similarly to the
Schaller et al. (1992) $20\msun$ one. It ascends the asymptotic
giant branch with highly convective envelope near the end of
evolution. PRH show in their Fig.~1 calculations performed without
wind loss for a large range of ZAMS masses
and they exhibit this behaviour.

We base our evolution on our Eq.~(\ref{eq1}) having established
in LBW that it is consistent with present observational data, if the
product $\lambda\alpha_{ce}$ in the Webbink (1984) common envelope
evolution is set equal to 0.2. Possible deviations from our
choice can be found from Fig.~3 of PRH for values of 0.5 and 0.08,
2.5 times greater and 2.5 times smaller than ours.

The chief difference in LBW and PRH is in the treatment of the
AML (angular momentum loss) SXTs. In the spiral in these
overfill their Roche Lobes with $\lambda\alpha_{ce}=0.2$,
presumably even more so with $\lambda\alpha_{ce}=0.08$.
In the donor mass considered ($M_d>0.7\msun$) of LBW the donor 
did not overfill its Roche Lobe by much.\footnote{
    And with slightly larger $\lambda\alpha_{ce}$ it would not
    overfill the Roche Lobe at all.
    }
It was assumed that the system adjusted
itself quickly by transfer of a small amount of mass to the He star,
which widened the orbit until the donor filled its Roche Lobe
exactly. In fact we now believe it more likely
that the overfill mass is expelled in the common envelope evolution.
Meyer \& Meyer-Hofmeister (1979) suggest that the common envelope
is not expelled until the separation of the inner cores (He-star and
donor) has become so small that the dense layers of the donor are
finally affected by the tidal interaction. ``At this point in the evolution
a large amount of mass is rather suddenly released from the main
sequence star into the common envelope and the neighborhood of the
degenerate binary companion."

PRH check whether the secondary star overfills its Roche Lobe.
If so they ``assume that the secondary merges with the core and do not
follow the binary further." This is presumably the reason that
they are unable to evolve the AMLs with their smaller
$\lambda=0.08$, 
in which case their
$a_f$ following common envelope evolution would be 2.5 times
smaller than ours. We believe that the great regularity in the
8 AMLs supports our $\lambda\alpha_{ce} \sim 0.2$.

One cannot pin down $\lambda$ separately from this combination, however,
Dewi \& Tauris (2001) suggest a lower limit of $\lambda=0.2$, whereas
for deeply convective giants and no wind loss
$\lambda\sim 1$ as discussed in our Appendix.
We can only guess that $\lambda\sim 0.5$, in the middle of the allowed
interval. This would give $\alpha_{ce}\sim 0.4$, saying that the
material released in the common envelope evolution has a kinetic
energy $2.5$ times the averate kinetic energy it had in the initial
giant. (Even though some of the released matter comes from the donor
through tidal interaction, most of it must come from the envelope
of the giant.) Most of this kinetic energy comes when the tidal
interaction has cut strongly into the donor.

In the best calculation of common envelope evolution to date,
Rasio \& Livio (1996) say that perhaps their most significant
new result is that during the dynamical phase of CE evolution,
a corotating region of gas is established near the central
binary. This is done through a combination of spiral shock
waves and gravitational torques that can transfer angular momentum
from the binary orbit to the gas. The corotating region has
the shape of an oblate spheroid encasing the binary (i.e., the
corotating gas is encased in the orbital phase).
The assumption that rigid rotation is tidally enforced in a core
surrounding the inner binary was already made by Meyer \& Meyer-Hofmeister
(1979). Rasio \& Livio (1996) did not carry their calculation beyond this
stage.

LBW assumed that at least the outer part of the He star is isochronous
with the donor at the time of the explosion forming the black hole. In
fact, the giant will probably have been brought into common
(rigid body) rotation by the dynamo process proposed by Spruit (2001)
and even more so by the magnetic field modeling suggested by Spruit
\& Phinney (1998). In general this rotation will be about an axis
different from that established later in the common envelope evolution.
The assumption that rigid rotation is tidally enforced in a core
surrounding the inner binary was already made by Meyer \&
Meyer-Hofmeister (1979). However, the common envelope time is short,
the dynamical time of years, so it would be expected to bring
only the outer part of the He core into synchronism.
The remaining He-star burning following common envelope evolution
of $\sim 100$ years seems too short to effect synchronization by itself.
However, in this time the inner core can pull away from the outer
He star (Spruit \& Phinney). Thus, we believe that sufficient differential
rotation will be achieved to allow the center of the He star to fall into
a black hole and the surrounding part into an accretion disk.

Wilson (1989) has examined synchronism in Algol systems. Out of 33
systems, about 2/3 show synchronism for periods less than two weeks.

With a $\lambda$-parameter of 0.5 PRH find a formation rate for
binaries with $m_d <2\msun$, the limit for donors which remain
in main sequence, of $7\times 10^{-7}$ yr$^{-1}$. They normalize
to a supernova rate of 1 per century. We use a rate of 3 per 
century, giving a formation rate of
$2.1\times 10^{-6}$ yr$^{-1}$. Although these live a Hubble time,
we can observe these for $10^9$ yrs (Lee \& Brown 2002) so this
would give 2100 presently observable in the Galaxy, of the same
general number as Wijers' estimate of 3000 (Wijers 1996).

PRH give $\gsim 10^{-5}$ as the formation rate for binaries with
$m_d<15\msun$. We wish to take an upper limit of at least $20\msun$,
to include \astrobj{Cyg X-1}, and to multiply their formation rate by our 3
normalization factor. We find them a formation rate of $\sim 5\times
10^{-5}$ yr$^{-1}$, a factor $\sim 4$ less than the estimate of 
Lee \& Brown (2002), but as noted there, a realistic number for those
which could be relics of GRBs might be as low as 
$\sim 10^{-5}$~galaxy$^{-1}$~yr$^{-1}$. In other words, the progenitor
binaries of the black-hole binaries are sufficient to also be
progenitors of GRBs.

%------------------------------------------------------------------
\section{Discussion}
\label{disc}

The work of LBW (Lee, Brown, \& Wijers 2002)
has been amalgamated with that of PRH (Podsiadlowski et al. 2002).
Both papers agree that common envelope evolution must come following
helium core burning; i.e., be Case C. Our work is based on the nearly
linear relationship Eq.~(\ref{eq1}) between separation following
common envelope evolution and companion (donor) mass, which is spelled out
in our Appendix. By choosing the common envelope parameter
$\lambda\alpha_{ce}=0.2$, we are able to evolve those binaries
with K and M-star companions, which we believe to be the success
of LBW and our earlier works.

The PRH evolutions clarify that all of the binaries with the possible
exception of \astrobj{Cyg X-1} could have made Roche contact in main
sequence evolution of the companion, as in the earlier work by
Beer \& Podsiadlowski (2002). This makes it possible to evolve
all of the binaries with sub-Eddington rate of mass transfer (although
we do not believe this to be necessary in the case of black holes).

LBW and PRH agree that the present evolutionary tracks of supergiants
from ZAMS masses $\sim 20-30\msun$ and possibly greater, must be
changed so as to allow Case C mass transfer. An example of how to
do this was constructed (by hand) in LBW.

We would like to point out that there are uncertainties due to the
radius evolution of massive stars and the parameterization of mixing, etc.
In particular, stars that use the LeDoux criterion or a small amount of
semiconvection burn helium as red supergiants, while those with
Schwarzschild or a lot of semiconvection stay blue much of the time.
Rotationally induced mixing also plays a role as in Langer and Maeder (1995).

%%%%%%%%%%%%%%%%%%%%%%%%%%%%%%%%%%%%%%%%%%%%%%%%%%%%%%%%%%%%%%%%%%%%%%%%%%%%%%%%
\section*{Acknowledgments}

GEB is supported by the U.S. Department of Energy under grant
DE-FG02-88ER40388. 
CHL is supported by Korea Research Foundation
Grant (KRF-2002-070-C00027).

\appendix
\section{Common Envelope Efficiency}
\label{efficiency}

The binding energy of the envelope is parameterized by
$\lambda$ (Webbink 1984),
\be
E_{\rm env} = -\frac{G M_{\rm giant} M_{\rm env}}{\lambda R},
\ee
where $M_{env}$ is the mass of the envelope in the giant star.
During the common envelope evolution, this binding can be compensated
by the total change in binary orbital energy with efficiency $\alpha_{ce}$
\be
E_{\rm env} = \alpha_{ce} \left(-\frac{G M_{\rm He} M_{\rm d}}{ 2 a_f}
+\frac{G M_{\rm giant} M_{\rm d}}{ 2 a_i}\right),
\label{eqA2}
\ee
where $a_{i,f}$ are the initial and final orbital separations and
$M_{\rm comp}$ is the donor companion mass, and one can neglect last term
with $a_i$. Hence, $a_f$ can be expressed as
\be
a_f &\approx & 
\lambda\alpha_{ce}  M_{\rm d} R \frac{M_{\rm He}}{2 M M_{\rm env}}
\nonumber\\
&\approx& 0.04 \; \lambda\alpha_{ce} \frac{M_d}{\msun} 
\left(\frac{M_{\rm giant}}{\msun}\right)^{-0.55} R
\; \times \frac{M_{\rm giant}}{M_{\rm env}}
\label{eqA3}
\ee
where we have used $M_{\rm He}=0.08 (M/\msun)^{1.45} \msun$.
The final factor $M_{\rm giant}/M_{\rm env}\simeq 3/2$ to within
5\% over our range of ZAMS masses $20-30\msun$.
Note that only the combination $\lambda\alpha_{ce}$ is relevant.
In LBW in the
commonly used Webbink formalism we found, analyzing the same orbits,
$\lambda\alpha_{ce}=0.2$. Here $\lambda$ is a parameter taking
into account the polytropic structure of the giant.
For deeply convective giants without mass loss, Brown et al. (2001b)
found $\lambda\simeq 7/6$ in their Appendix C.

\begin{figure}
\centerline{\epsfig{file=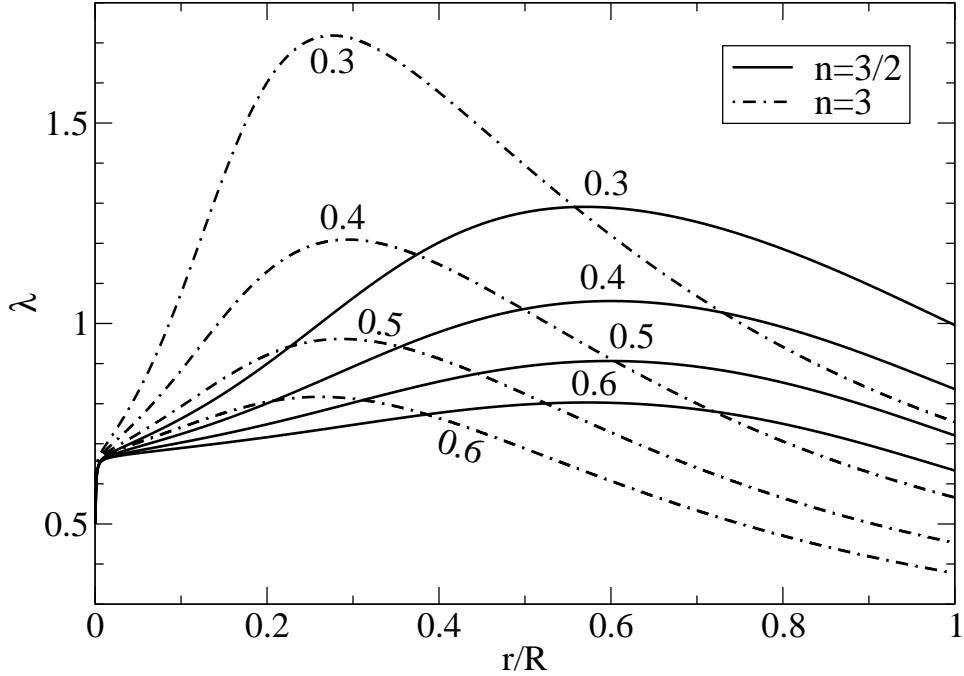,height=4.5in}}
%\centerline{\epsfig{file=x.eps,height=3.5in}}
\caption{Estimated $\lambda$ parameter of the mass losing star
with convective (n=3/2) and 
radiative (n=3) envelopes.
The numbers in the plot are the
mass ratios of the core to the total star mass, $x$.
Note that the He core mass fraction of ZAMS $20-40\msun$
star is $0.3 < x <0.4$ with assumed formula
$M_{\rm He}=0.08 M^{0.45}$. However, $x$ will increasing
while losing mass, and larger $x$ is more relevant for the
later stages of the common envelope evolution.
Since the star changes structure while losing mass and
the size of core is not negligible for small $r$,
the physically relevant region will be near $r/R\sim 1$. 
}
\label{fig_lambda}
\end{figure}

In order to get the estimated values of $\lambda$ of the mass losing
star (Dewi \& Tauris 2001), we calculated the binding energy parameter 
$\lambda$ for the convective (polytropic index $n=3/2$) and radiative 
($n=3$) envelopes in Fig.~\ref{fig_lambda}.
In this figure, we considered various values of fractional
core mass. Note that 
the simple formula used in our calculation $M_{\rm He}=0.08 M^{1.45}$
gives initial values of $x$ to be $0.3-0.4$
for the ZAMS mass range $20-40\msun$. However, the fractional core mass
will increase during the common envelope evolution.
In the estimation of $\lambda$ of the mass losing star in Fig.~\ref{fig_lambda},
we assumed that i) the size of core is negligible compared to the radius
of the star $R$, ii) the structure of the original star doesn't change when
we remove part of envelope outside of radius $r$. So, the plotted
$\lambda$ is the binding energy parameter of the remaining star inside
radius $r$.

Since the star changes structure while losing mass and
the core size is not negligible for small radius,
the physically relevant region will be large radius
near $r/R\sim 1$. For the
fully convective star with $x=0.4$, the binding energy parameter
$\lambda\sim 0.75$. Since the fractional mass ratio $x$ increases
during the common envelope evolution and the
outer radiative ($n=3$) part of the envelope will decrease
$\lambda$, lower average $\lambda\sim 0.5-0.7$ is a reasonable approximation
for the common envelope evolution. 
So the efficiency parameter becomes $\alpha_{ce}\sim 0.3-0.4$ to
have $\lambda\alpha_{ce}=0.2$ found in LBW.

%%%%%%%%%%%%%%%%%%%%%%%%%%%%%%%%%%%%%%%%%%%%%%%%%%%%%%%%%%%%%%%%%%%%%%%%%%%%%%%%
%   references
%%%%%%%%%%%%%%%%%%%%%%%%%%%%%%%%%%%%%%%%%%%%%%%%%%%%%%%%%%%%%%%%%%%%%%%%%%%%%%%%
%\bibliographystyle{astroshort}
%\bibliography{xmymoshort,x65,x70,x75,x80,x85,x90,x95,x00,x05}

\begin{thebibliography}{}

\bibitem[\protect\astroncite{Amati et~al.}{2000}]{Amati00}
Amati, L., et~al.\ 2000,
\newblock Science 290, 953

\bibitem{Beer2001}
Beer, M. and Podsiadlowski, Ph. 2002, \MNRAS {\bf 331}, 351.

\bibitem[\protect\astroncite{}{Brown et al. 2001a}]{Brown01A}
Brown, G.E., Heger, A., Langer, N., Lee, C.-H.,  Wellstein, S.,
and  Bethe, H.A. 2001a, \newast {\bf 6}, 457.

\bibitem{Brown01B}
Brown, G.E., Lee, C.-H., Portegies Zwart, S.F., and Bethe, H.A.
2001b, \ApJ {\bf 547}, 345.

\bibitem{Brown01C}
Brown, G.E., Lee, C.-H., and Tauris, T. 2001c, \newast {\bf 6}, 331.

\bibitem{DeKool87}
De Kool, M.,  van den Heuvel, E.P.J., and  Pylyser, E. 1987, 
\AnA {\bf 183}, 47.

\bibitem{Dewi}
Dewi, J. D. M. and Tauris, T. M. 2001, Proc. of ``Evolution
of Binary and Multiple Star Systems", ASP Conference Series, Vol. 229, 
eds. Ph. Podsiadlowski, S. Rappaport, A. R. King, F. D'Antona,
and L. Burderi, p. 255.

\bibitem{Fryer}
Fryer, C.L., Heger, A., Langer, N., and Wellstein, S. 2002, \ApJ
{\bf 578}, 335.


\bibitem{Garcia97}
Garcia, M.R., McClintock, J.E.,  Narayan, R., and  Callanan, R.J. 1997,
Proceedings of the 13th North American Workshop on CVs, eds. S. Howell,
E. Kuulkers, C. Woodward (San Francisco: ASP) 506 (1997);
astro-ph/9708149.

\bibitem{Langer1995}
Langer, N. and Maeder, A. 1995, \AnA, {\bf 295}, 685.

\bibitem{Lee2002b}
Lee, C.-H. and Brown, G.E. 2003, Int. J. of Mod. Phys. {\bf A 18}, 527.

\bibitem[\protect\astroncite{Woosley and Weaver}{1995}]{Lee2002}
Lee, C.-H., Brown, G.E., and Wijers, R.A.M.J. 2002, \ApJ {\bf 575}, 996 (LBW).

\bibitem{meyer}
Meyer, F. and Meyer-Hofmeister, E. 1979, \AnA {\bf 78}, 167.

\bibitem{Nelemans01}
Nelemans, G. and van den Heuvel, E.P.J. 2001, A\&A, {\bf 376}, 950.

\bibitem{podsiad}
Podsiadlowski, Ph., Rappaport, S., and Han, Z. 2003,  MNRAS {\bf 341}, 385
 (PRH).
\bibitem{Simon97}
Portegies Zwart, S.F., Verbunt, F., and Ergma, E. 1997, \AnA {\bf 321} 207.

\bibitem{Rasio}
Rasio, F.A. and Livio, M. 1996, \ApJ {\bf 471}, 366.

\bibitem[\protect\astroncite{}{Schaller et al. 1992}]{Schaller92}
Schaller, G., Schaerer, D., Meynet, G., and  Maeder, A. 1992,
\AnAS {\bf 96}, 269.

\bibitem{tauris99}
Tauris, T.M. and Savonije, G.J. 1999, A\&A, {\bf 350}, 928.

\bibitem{Webbink}
Webbink, R.F. 1984, \ApJ {\bf 277}, 355.

\bibitem{wijers96}
Wijers, R.A.M.J. 1996, Evolutionary Processes in Binary Stars,
Eds. R.A.M.J. Wijers et al. Kluwer, Dordrecht, pp. 327-344.

\bibitem{Woosley}
Woosley, S.E., Langer, N., and Weaver, T.A. 1995, \ApJ {\bf 448} 315.

\bibitem[\protect\astroncite{Woosley and Weaver}{1995}]{Woosley95B}
Woosley, S.~E. \& Weaver, T.~A. 1995,
\newblock ApJS {\bf 101}, 181

\end{thebibliography}

\def\ApJ{{\it Astrophysical Journal}$\;$}
\def\ApJL{{\it Astrophysical Journal Letters}$\;$}
\def\AJ{{\it Astronomical Journal}$\;$}
\def\newast{{\it New Astronomy}$\;$}
\def\ARAnA{{\it Annual Reviews of Astronomy and Astrophysics}$\;$}
\def\AnA{{\it Astron. and Astrophys.}$\;$}
\def\AnAS{{\it Astron. and Astrophys. Suppl.}$\;$}
\def\MNRAS{{\it Mon. Not. of Royal. Astron. Soc.}$\;$}
\def\PASP{{\it Pub. of the Astron. Soc. of the Pac.}$\;$}

\end{document}